\documentclass[draft]{agujournal2019}
\usepackage{url} 
\usepackage{lineno}
\usepackage[inline]{trackchanges} 
\usepackage{soul}
\usepackage{amsmath}
\usepackage{graphicx}
\usepackage{dcolumn}
\usepackage{bm}
\usepackage{ragged2e}

\usepackage{latexsym}
\usepackage{lipsum}
\usepackage{epsf}
\usepackage{float}

\usepackage{epsfig}
\usepackage{subfigure}

\usepackage{xcolor}
\usepackage{tikz}
\usetikzlibrary{shapes}
\definecolor{blue}{RGB}{0,112,192}
\definecolor{lightblue}{RGB}{0,176,240}
\definecolor{green}{RGB}{0,176,80}
\definecolor{yellow}{RGB}{255,255,0}
\definecolor{orange}{RGB}{255,192,0}
\definecolor{red}{RGB}{255,0,0}
\definecolor{darkred}{RGB}{118,0,0}
\definecolor{purple}{RGB}{208,0,154}

\newcommand*{\tikzrect}[2]{%
   \setbox0=\hbox{\strut}%
   \begin{tikzpicture}
     \useasboundingbox (-.25em,0) rectangle (.25em,\ht0);
     \filldraw[draw=#1,fill=#2] (-0.25em,0.05em) rectangle (.25em,0.55em);
   \end{tikzpicture}%
}

\newcommand{\tikztri}[2]{\tikz{\node[draw=#1,fill=#2,isosceles triangle,isosceles triangle stretches,shape border rotate=90,minimum width=0.2cm,minimum height=0.2cm,inner sep=0pt] at (0,0) {};}}

\draftfalse

\journalname{Geophysical Research Letters}

\begin{document}

%
%

\title{Finite-Size Analysis of the Collapse of Dry Granular Columns}

%
%




\authors{Teng Man\affil{1}, Herbert E. Huppert\affil{2}, Ling Li\affil{1}, Sergio A. Galindo-Torres\affil{1}}

\affiliation{1}{Key Laboratory of Coastal Environment and Resources of Zhejiang Province (KLaCER), School of Engineering, Westlake University, 18 Shilongshan Street, Hangzhou, Zhejiang 310024, China}
\affiliation{2}{Institute of Theoretical Geophysics, King's College, University of Cambridge, King's Parade, Cambridge CB2 1ST, United Kingdom}





\correspondingauthor{S.A. Galindo-Torres}{s.torres@westlake.edu.cn}




\begin{keypoints}
\item Granular column collapses show a significant size effect with respect to run-out distances and phase transitions;
\item The size effect can be further associated with the strong force networks and early-stage characteristic length scales;
\item We draw similarities between granular column collapses and dome-collapse pyroclastic flows.
\end{keypoints}

%
%

%
%

\justifying  
\begin{abstract}
In this letter, we focus on the size effect of granular column collapses, which are potentially connected to the dynamics of complex geophysical flows, even if the link between microscopic structures of granular assemblies and their macroscopic behaviors is still not well understood. Using the sphero-polyhedral discrete element method (DEM), we show that the column radius/grain size ratio has a strong influence on the collapse behavior. A finite-size analysis, which is inspired by a phase transition around an inflection point, is performed to obtain a general scaling equation with critical exponents for run-out distances. We further link the size effect with the strong force network and formalize a correlation length scale that exponentially scales with the effective aspect ratio. Such a scaling solution shows similarities with the percolation problem of two-dimensional random networks and can be extended to other similar natural and engineering systems.
\end{abstract}

\section*{Plain Language Summary}
We aim to understand the size effect in granular column collapses. Geophysical hazards, such as landslides, debris flows, pyroclastic flows, and rock avalanches, occur frequently worldwide, and it is difficult to predict their initiation and dynamical behavior. Many geophysical flows can be considered as granular materials, where interaction between particles matters. However, it is difficult to upscale a grain-scale particle interaction to a large-scale predicting tool to capture the behavior of many complex granular flows. This paper starts from the collapse of granular columns, a fundamental problem often used to investigate the mobility of granular materials, and carefully navigates to examine the run-out behaviors of columns with different sizes, so that we can obtain a better picture of how different system sizes lead to different collapsing scenarios. We further analyzed the characteristic length scale associated with the collapse of granular columns, and the strong force network presenting at the beginning of the collapse, to understand better what drives the size effect of granular systems, and what we can gain from this research to predict the behavior of geophysical flows.

\section{Introduction}
Granular materials are omnipresent in natural and engineering systems, and the physics and mechanics of them are crucial for understanding geophysical flows, natural hazards, food processing, chemical engineering, and pharmaceutical engineering \cite{Lube2019GenerationOA,guyon2020built}. Granular materials can behave like a solid, a liquid, or a gas in different circumstances \cite{midi2004dense}, which increases the difficulty in capturing their macroscopic behavior. Besides, collective structures may form inside a granular system, and the existence and the size of such collective structures [such as bridging \cite{mehta2007granular}, granular agglomerates \cite{Vo2020evolution}, and contact networks \cite{Zhang2014force}] will introduce a strong size effect, which further increases the complexity of the problem. In recent decades, breakthroughs have been made to understand the basic governing principles, especially the constitutive relationships, of granular materials \cite{midi2004dense,trulsson2012transition,Vo2020additive,man2021granular}, where the behavior of granular materials or granular-fluid mixtures are considered to be described by dimensionless numbers expressed as the ratio between dominating stresses. Even though geophysical flows are rarely dry, investigating dry granular materials is still considered as an important start for more complex systems.

Due to the similarity and potential links between the collapse of granular columns and gravity-driven geophysical flows, previous research investigated the collapse of granular columns to analyze the post-failure behavior of granular systems \cite{thompson2007granular,lacaze2009axisymmetric}. \citeA{lube2004axisymmetric} and \citeA{lajeunesse2005granular} independently determined relationships for both the normalized run-out distance $\mathcal{R}=(R_{\infty}-R_i)/R_i$ (where $R_{\infty}$ is the final radius of the granular pile, and $R_i$ the initial radius of the granular column), and the halt time of a collapsed granular column, which both scale with the initial aspect ratio, $\alpha = H_i/R_i$ (where $H_i$ is the initial height of the column), a parameter drawn out of dimensional analysis. \citeA{zenit2005computer,staron2005study,staron2007spreading,lacaze2009axisymmetric} studied either 2D or axisymmetric granular column collapses using the discrete element method (DEM), and explored the influence of inter-particle properties, energy dissipation, and inertial effects on the run-out and deposition of granular columns. \citeA{Farin2019RelationsBT} even linked the characteristics of granular column collapses to seismic signals. \citeA{warnett2014scalings} and  \citeA{Cabrera2019Granular} studied the collapse of granular columns with experiments and simulations, and argued that the relative size of a granular column, $R_i/d$, where $d$ is the average particle diameter, has strong influence on the run-out distance of a collapsed granular column, and to avoid significant size effects, $R_i/d$ must be larger than 75 for short columns and larger than 50 for tall columns. 

Based on dimensional analysis and simulations, we investigated the collapse of axisymmetric granular columns and their resulting deposition with a wide range of inter-granular and particle/boundary frictional coefficients \cite{man2020universality}, and concluded that the normalized run-out distance, $\mathcal{R}$, scales with an effective aspect ratio, 
\begin{linenomath*}
\begin{equation}
\begin{split}
    \alpha_{\textrm{eff}} = \left(\mu_w+\beta\mu_p\right)^{-1/2}\left(H_i/R_i\right)\ ,
\end{split}
\end{equation}
\end{linenomath*}
where $\mu_w$ is the frictional coefficient between particles and the boundary, $\mu_p$ is the inter-particle frictional coefficient, and $\beta$ can be seen as the ratio between contributions from inter-particle frictions and those from particle/boundary frictions. They found that the collapse of granular columns can be classified into three different regimes: quasi-static, inertial, and liquid-like. The effective aspect ratio, which was used to describe the run-out behavior, was obtained from dimensional analysis and was shown to represent the ratio between the inertial stress and the frictional stress or, equivalently, the ratio between the kinetic energy and dissipated energy during the collapse of granular columns. In our study, as we change the frictional coefficient, the transition point, where the slope of $\mathcal{R}(\alpha)$ curve changes on a log-log coordinate system, also changes accordingly. Moreover, after using $\alpha_{\textrm{eff}}$ to re-scale the data, this transition point in $\mathcal{R}-\alpha_{\textrm{eff}}$ space becomes universal.

In this paper, after introducing the simulation set-up and the contact law in the DEM, we further conduct a systematic study of granular column collapses with the sphero-polyhedral DEM to investigate the influence of relative column size on the run-out distance to obtain a scaling solution to describe granular column collapses with different sizes. This can not only further explore the physics of granular column collapses but also open a window for linking the behavior of granular columns to the mobility of geophysical flows. Inspired by the work of \citeA{Roche2002ExperimentsOD,Roche2008ExperimentalOO,bougouin2021}, where they presented physical connections between granular dam breaks and pyroclastic flows, to strengthen the tie between granular column collapses and geophysical flows, we present an application of our results to various rock falls around the Soufriere Hills, Montserrat as a result of volcanic eruptions and the resulting pyroclastic flows and granular surges in Section \ref{sec:application} and Figure \ref{discussion}. Finally, we present our concluding remarks.

\section{Simulations}\label{sec:simu}

\subsection{Discrete element method and simulation setup}


We perform DEM simulations of the granular column collapses with Voronoi-based sphero-polyhedra particles\cite{pournin2005generalization,galindo2010molecular}. The sphero-polyhedra technique allows for an easy and efficient definition of contact laws between particles. This is due to the smoothing of the edges of all geometric features by circles (in 2D) or spheres (in 3D). We note that, in our simulation, three types of contacts (vertex-vertex contact, edge-edge contact, and vertex-face contact) are considered. For these types of contacts, we implement a Hookean contact model with energy dissipation to calculate the interactions between particles. The DEM model used in this work has been thoroughly validated with experiments as shown in the supplemental material, where the shape of particles was accurately captured and the frictional coefficients of both inter-particle contact and particle/boundary contact were independently measured. 

\begin{figure}[!h]
  \centering
  \includegraphics[scale = 0.30]{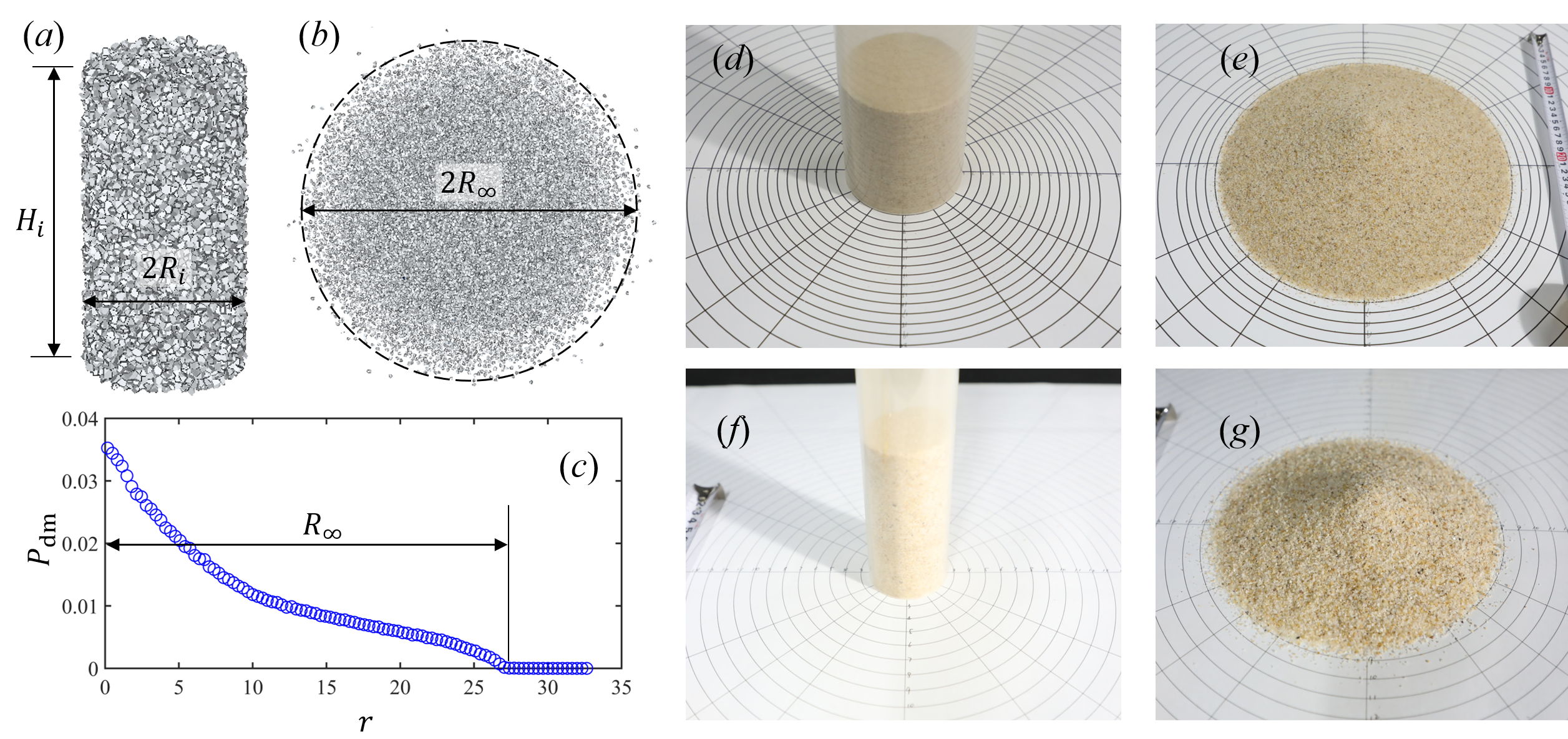}
  \caption{Simulation set-up. (a) the initial state of the granular column; (b) the final deposition of the collapsed column in a 3-D view; (c) shows the method for measuring the run-out distance of a collapsed granular column. The $x-$axis is the radial position $r$, and the $y-$axis is the percentage of number of particles, $P_{\textrm{dm}}$, located within $(r-\Delta r/2, r+\Delta r/2)$ divided by the radial position. We determine the final radius of a collapsed granular column when $P_{\textrm{dm}}(r) \leq 5\%P_{\textrm{dm}}^{\textrm{max}}$; (d) Initial state and (e) final state of granular column with $R_i = 5.7$ cm and $H_i = 13$ cm; (f) Initial state and (g) final state of granular column with $R_i = 2.3$ cm and $H_i = 10$ cm.}
  \label{setup}
\end{figure}

In the simulation, we create particles in a certain cylindrical domain with initial height, $H_i$, and initial radius, $R_i$ [Fig. \ref{setup}(a)]. 20\% of the particles are randomly chosen and removed from the simulation to create a granular packing with an initial solid fraction of $\phi_s = $0.8. The average particle size $d$ is 0.2 cm. The coefficient of restitution of particle collisions is 0.1. The material properties are set to be the same as that of quartz sand (particle density 2.65 g/cm$^3$). Using 3D Voronoi algorithm to create sphero-polyhedral particles, we can make sure that, for each particle, the length scales in different directions are kept approximately the same ( as shown in the supplemental material). This paper focuses entirely on simulations with granular columns of circular cross-sections. More general cross-sections are currently under investigation and will be presented in a subsequent publication. The particle/boundary frictional coefficient $\mu_w$ is 0.4, while the inter-particle frictional coefficients $\mu_p$ are 0.2, 0.4, and 0.6.

We treat the relative column size ($R_i/d =$ 2, 2.5, 3.75, 5, 7.5, 10, 12.5, 15, 17.5, 20, 30) as a key parameter. Within one set of simulations, with the same $R_i/d$, we varied the initial height $H_i$ and the inter-particle frictional coefficient $\mu_p$ to obtain the general collapse behavior with a wide range of initial conditions. After the initial Voronoi-based particle packing is generated, we delete the confining tube in the simulation to release all the particles under gravity. After releasing particles to the horizontal plane, the granular material will form a pile of loosely packed grains [Fig. \ref{setup}(b)], with final packing radius, $R_{\infty}$. Thus, the normalized run-out distance, $\mathcal{R}=(R_\infty - R_i)/R_i$, could be obtained. With Voronoi-based spheropolyhedra particles [Fig. \ref{setup}(a) and (b)], we implement the Hookean contact model with an energy dissipation term to calculate the interactions among contacting particles \cite{galindo2010molecular}. In this paper, the contact parameters are the same as previous papers \cite{galindo2013coupled,man2020universality}.

\subsection{\label{2_2}Determination of the run-out distance}

In simulations, the measurement of the final packing radius is complicated due to the lack of a clear mathematical definition of the run-out distance. In cases with small particle/boundary and inter-particle frictional coefficients but large initial aspect ratios, the spread of particles is far-reaching and leads to sparse (single layer) coverage of the area, especially at the front edge. In these cases, it is difficult to determine the edge/boundary and hence the final run-out distance. Thus, we measured the final radius with a histogram of particle distribution for each simulation. Fig. \ref{setup}(c) gives an example of how we measured the run-out distance.

In Fig. \ref{setup}(c), the $x$-axis is the radial position $r$, and the $y$-axis is the percentage of number of particles located within $(r-\Delta r/2, r+\Delta r/2)$ divided by the radial position, $P_{\textrm{dm}}(r) = (1/r)[N(r-\Delta r/2, r+\Delta r/2)/(\Sigma_rN)]$, where $\Delta r$ is the bin width of the histogram, $N(r-\Delta r/2, r+\Delta r/2)$ is the number of particles located between $r-\Delta r/2$ and $r+\Delta r/2$, and $\Sigma_rN$ is the total number of particles in one simulation. FIG. \ref{setup}(c) shows the normalized particle number distribution, in this case, deposition morphology, of a simulation with $\mu_w=0.4$, $\mu_p=0.4$, $R_i = 4$ cm, and $H_i=32$ cm. It shows that most particles locate within $r \leq 27$ cm. Thus, we take the final deposition radius as $R_{\infty} = 27$ cm. After determining the final run-out distance of the collapsed granular column, we obtain the normalized run-out distance, $\mathcal{R}$, the initial aspect ratio, $\alpha$, and the effective aspect ratio \cite{man2020universality}, $\alpha_{\textrm{eff}}$, according to
\begin{linenomath*}
\begin{subequations} \label{eq:normalized_runout}
\begin{align}
    \mathcal{R} &= \left(R_{\infty} - R_i\right)/{R_i}\ ,\ \ \alpha = H_i/R_i\ ,\\
    \alpha_{\textrm{eff}} &= \alpha\left(\mu_w + \beta\mu_p\right)^{-1/2}\ .
\end{align}
\end{subequations}
\end{linenomath*}

\section{Experiments}\label{sec:exp}

\citeA{warnett2014scalings} and \citeA{Cabrera2019Granular} examined the size effect in the granular column collapses with experiments. In this work, we present two sets of experiments to show that changing sizes could influence both the $\mathcal{R}-\alpha$ relationship and the transition point. We perform column collapses of sand particles with grain sizes ranging from 1 mm to 2 mm (the average particle diameter $\bar{d}\approx 1.5$ mm). The experimental setup is the same as that in the simulations. The column diameters are 4.6 cm and 11.4 cm, respectively, resulting in relative column sizes $R_i/d \approx$ 15.3 and 38 [Fig. \ref{setup}(d)-(g)]. We vary the initial height of the column so that the initial aspect ratio varies from $\approx 0.3$ to $\approx 20$. After randomly dropping sand grains into the cylindrical tube, we measure the initial height, $H_i$, of the column, and then quickly lift the tube upward to let the granular column collapse. We measure the final radius of the sand pile in four different directions and take their averages as the final run-out distance, $R_{\infty}$. Then, the relationship between the initial aspect ratio, $\alpha$, and the normalized run-out distance, $\mathcal{R}$, can be obtained accordingly.

\begin{figure}[!ht]
  \centering
  \includegraphics[scale = 0.38]{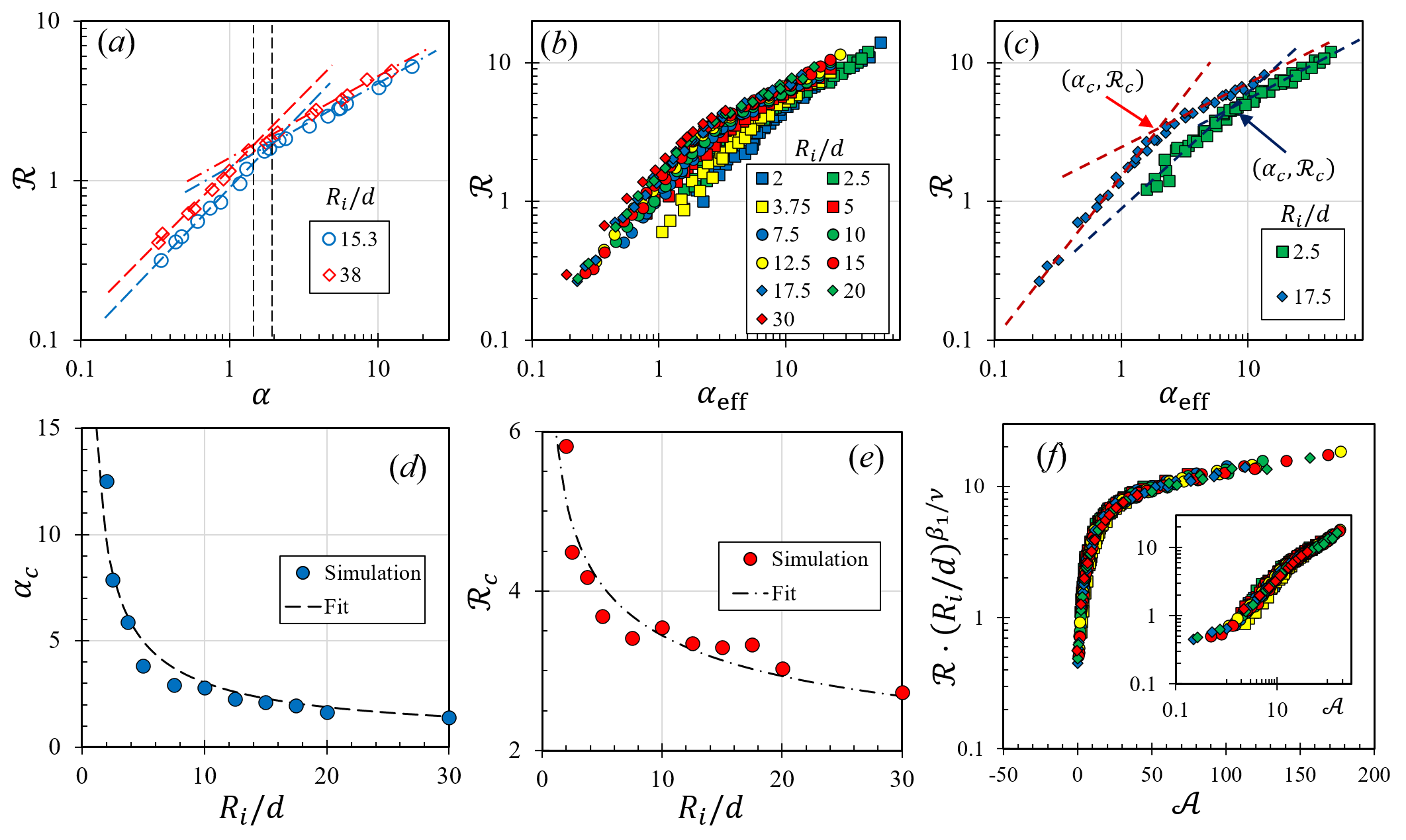}
  \caption{(a) The relationship between the normalized run-out distance, $\mathcal{R}=(R_{\infty}-R_i)/R_i$, and the initial aspect ratio, $\alpha = H_i/R_i$, of the experiments. The blue dashed line and the blue dash-dot line represent fitting curves before and after \add{the} transition point, when $R_i/d=15.3$. The green dashed line and the green dash-dot line denote the fitting curves before and after \remove{its}\add{the} transition point, when $R_i/d=38$. The black dashed line marks the transition aspect ratio of $R_i/d=15.3$, while the black dash-dot line represents the transition aspect ration of $R_i/d=38$; (b) $\mathcal{R}$ against the effective aspect ratio, $\alpha_{\textrm{eff}}=\alpha\sqrt{1/(\mu_w+2\mu_p)}$ for all the column sizes in the simulations; (c) the relationship between $\mathcal{R}$ and $\alpha_{\textrm{eff}}$ with relative column size $R_i/d =$ 2.5 and 17.5, respectively, to show that the critical inflection point $(\alpha_c,\ \mathcal{R}_c)$ changes as we change the relative size of the column; (d) \remove{shows} the relationship between transitional aspect ratio, $\alpha_c$, and the relative column size, $R_i/d$. (e) \remove{plots} the transitional normalized run-out distance, $\mathcal{R}_c$ against $R_i/d$; (f) \remove{shows} the relationship between $\mathcal{R}(R_i/d)^{\beta_1/\nu}$ and $\mathcal{A}= (\alpha_{\textrm{eff}}-\alpha_{c\infty})(R_i/d)^{1/\nu}$, with $\nu = 1.39$ and $\beta_1 = 0.28$. The inset of the figure plots the same relationship in a log-log coordinate system.}
  \label{results}
\end{figure}

We plot the experimental relationship between the initial aspect ratio and normalized run-out distance in Fig. \ref{results}(a). For granular column collapses with both column sizes, as we increase the initial aspect ratio, the normalized run-out distance, $\mathcal{R}$ increases accordingly. We can then observe the clear transition point at $\alpha \approx 1.2$ for cases with $R_i = 5.7$ cm and at $\alpha \approx 2.0$ for cases with $R_i =2.3$. The experiments show that the transition point becomes smaller as we increase the relative column size, and generally, the normalized run-out distance of large columns is larger than that of small columns. The experiments show clear size effect for granular column collapses. It is an experimental proof that changing relative column sizes can not only influence the run-out distance but also affect the transition point on the $\mathcal{R}-\alpha$ relationship. We note that the friction between sand particles and the confining tube while pulling it up might also influence the experimental results, but the influence is not significant, and it does not affect the transition we observe in the $\mathcal{R}-\alpha$ relationship. In the supplemental material, we also present a set of experiments and corresponding simulations of the collapse of cubic particle assemblies. The agreement between experimental results and simulations shows that our DEM model could well capture the behavior of gravity-driven transient granular flows.

\section{\label{res_discus}Simulation Results and Discussions}

\subsection{Size effects of granular column collapses}

After simulating granular column collapses with various column sizes and frictional coefficients, we obtain the relationship between $\mathcal{R}$ and $\alpha$. Similar to what we have seen previously \cite{man2020universality}, for columns with the same relative size $R_i/d$, less friction leads to larger run-out distances. As we change the $x-$axis to the effective aspect ratio, $\alpha_{\textrm{eff}}$, normalized run-out distance with the same $R_i/d$ collapse onto one curve in $\mathcal{R}-\alpha_{\textrm{eff}}$ space [Fig. \ref{results}(b) and (c)]. The previous work indicates that the effective aspect ratio, which includes the influence of both initial aspect ratio and particle frictional properties, could describe the run-out behavior of the granular column collapse, and most importantly it shows that, for columns with the same relative column size, there is only one transition point for certain column size in the $\mathcal{R}-\alpha_{\textrm{eff}}$ relationship. The collapsing of the transition point of the $\mathcal{R}-\alpha_{\textrm{eff}}$ relationship indicates a possible phase transition related to the collapsing dynamics of granular columns.

However, with various relative column sizes, we see that, in both Fig. \ref{results}(b) and (c), simulation results, especially the transition point in each $\mathcal{R}-\alpha_{\textrm{eff}}$ relationship, varies as we change $R_i/d$. For granular columns with the same initial aspect ratio, granular columns with larger relative column size $R_i/d$ have longer run-out distances. In Fig. \ref{results}(b), we plot simulation results for all the system sizes, and simulation results with $R_i/d = 2.5$ and $R_i/d = 17.5$ in Fig. \ref{results}(c). The latter gives us a clearer picture of how changing relative system size could result in variations in the relationship between $\mathcal{R}$ and $\alpha_{\textrm{eff}}$ and variations in the transition point $(\alpha_c, \mathcal{R}_c)$. This shows that the collapse of the granular column has a significant size effect, which was also observed by \citeA{warnett2014scalings} and \citeA{Cabrera2019Granular}. However, no quantitative studies has been made to universally include different frictional coefficients and boundary conditions to describe the run-out behavior and few researchers pointed out the physical nature of the initial aspect ratio $\alpha$ other than obtaining it from dimensional analysis \cite{man2020universality}. 

Further, changing the relative column size also fundamentally influences the shape of $\mathcal{R}-\alpha_{\textrm{eff}}$. For simulations with the same $R_i/d$, a threshold $\alpha_c$ of $\alpha_{\textrm{eff}}$ exists to divide the $\mathcal{R}-\alpha_{\textrm{eff}}$ relationship into two groups [Fig. \ref{results}(b)]. When $\alpha_{\textrm{eff}} < \alpha_c$, $\mathcal{R}$ approximately scales with $\alpha_{\textrm{eff}}$ proportionally. When $\alpha_{\textrm{eff}} > \alpha_c$, $\mathcal{R}$ approximately scales with $(\alpha_{\textrm{eff}})^{0.5}$ with a rather sharp division between the two. Here, the transitional aspect ratio $\alpha_c$ and the corresponding transitional normalized run-out distance $\mathcal{R}_c$ can be seen as the critical aspect ratio and the critical run-out distance, respectively. Both $\alpha_c$ and $\mathcal{R}_c$ vary as we change the size of the granular column. For instance, in Fig. \ref{results}(c), the transition happens at $\alpha_c \approx 8, \mathcal{R}_c \approx 5$ when $R_i/d = 2.5$, and happens at $\alpha_c \approx 2, \mathcal{R}_c \approx 3.5$ when $R_i/d = 17.5$.


We show, in Fig. \ref{results}(b) and (c), that a critical transition point ($\alpha_c$ and the corresponding $\mathcal{R}_c$) exists and also varies with different relative size $R_i/d$. Thus, the $\mathcal{R}-\alpha_{\textrm{eff}}$ curve is dictated by the position of  $\alpha_c$ and $\mathcal{R}_c$ that
\begin{linenomath*}
\begin{equation} \label{eq_general}
    \begin{split}
        \mathcal{R} = f\left( \alpha_{\textrm{eff}}-\alpha_c, \mathcal{R}_c, R_i/d\right).
    \end{split}
\end{equation}
\end{linenomath*}
Both $\alpha_c$ and $\mathcal{R}_c$ decrease with \remove{an} increasing $R_i/d$. In Fig. \ref{results}(d) and (f), based on the relationship between $\alpha_{\textrm{eff}}$ and $\mathcal{R}$ in Fig. \ref{results}(b) and (c), we plot the relationship between $\alpha_c$ and $R_i/d$ and the relationship between $\mathcal{R}_c$ and $R_i/d$. Thus, we could write $\alpha_c$ and $\mathcal{R}_c$ as functions of the relative column size, where, when $R_i/d$ approaches infinity, both $\alpha_c$ and $\mathcal{R}_c$ converge to $\alpha_{c\infty}$ and $\mathcal{R}_{c\infty}$, as shown by
\begin{linenomath*}
\begin{subequations} \label{eq_transition}
    \begin{align}
        \alpha_c = \alpha_{c\infty} + a_1\left( {R_i}/{d}\right)^{b_1}\ , \\
        \mathcal{R}_c = \mathcal{R}_{c\infty} + a_2\left( {R_i}/{d}\right)^{b_2}\ ,
    \end{align}
\end{subequations}
\end{linenomath*}
where $\alpha_{c\infty} = 0.2$ and $\mathcal{R}_{c\infty} = 0.732$ are the fitted critical aspect ratio and the corresponding critical run-out distance, respectively, when the relative column size $R_i/d$ goes to infinity, and both parameters are fitted values. Also, $a_1 = 16$, $b_1 = -0.75$, $a_2 = 5.4$, and $b_2 = -0.3$ are fitting parameters. The fitted curves of $\alpha_c$ and $\mathcal{R}_c$ are the dashed curve in Fig. \ref{results}(d) and the dash-dot curve in Fig. \ref{results}(e).


The power-law decay of both $\alpha_c$ and $\mathcal{R}_c$ with respect to $R_i/d$ inspires us to perform a finite-size analysis of the run-out distance of the collapse of granular columns based on our working hypothesis of a potential phase transition. Since a general scaling analysis often works under thermodynamics limits, where the system size and its corresponding particle number is assumed to be infinity. Finite-size analysis then becomes a powerful tool for investigating real systems with finite sizes. As shown in Fig. \ref{results}(f), the normalized run-out distance, $\mathcal{R}$ indeed exhibits excellent finite-size scaling, suggesting that the transitional aspect ratio, $\alpha_c$, is critical. All the normalized run-out distance data collapse nicely onto a master curve in the form \cite{Torres2015scaling}
\begin{linenomath*}
\begin{equation} \label{eq_fss}
    \begin{split}
        \mathcal{R} = \left({R_i}/{d}\right)^{-\beta_1/\nu}\mathcal{F}_r\left[(\alpha_{\textrm{eff}}-\alpha_{c\infty})\left({R_i}/{d}\right)^{1/\nu}\right]\ ,
    \end{split}
\end{equation}
\end{linenomath*}
with the limiting scaling of the normalized run-out distance being $\mathcal{R}\sim \alpha_{\textrm{eff}} - \alpha_{c\infty}$, where $\nu = 1.39\pm 0.14$ and $\beta_1 = 0.28\pm 0.04$ are obtained to best collapse all the data. In Fig. \ref{results}(f), to better represent the results, we took $\mathcal{A}= (\alpha_{\textrm{eff}}-\alpha_{c\infty})(R_i/d)^{1/\nu}$ as the $x-$axis. Since the system is axisymmetric, the scaling solution for the collapse of granular materials shows similar phenomena in the scaling solutions for connectivity of two-dimensional continuous random networks \cite{Torres2015scaling}, where the scaling parameters $\nu = 1.33$ and $\beta = 0.1389$. This indicates that the flowing behavior of granular column collapse has strong correlations with the connectivity of grain contact networks. This also coincides with the suggestion made by \citeA{mehta2007granular} that the transport of grains takes place percolatively, especially in the context of avalanches.

\begin{figure}[!ht]
  \centering
  \includegraphics[scale = 0.35]{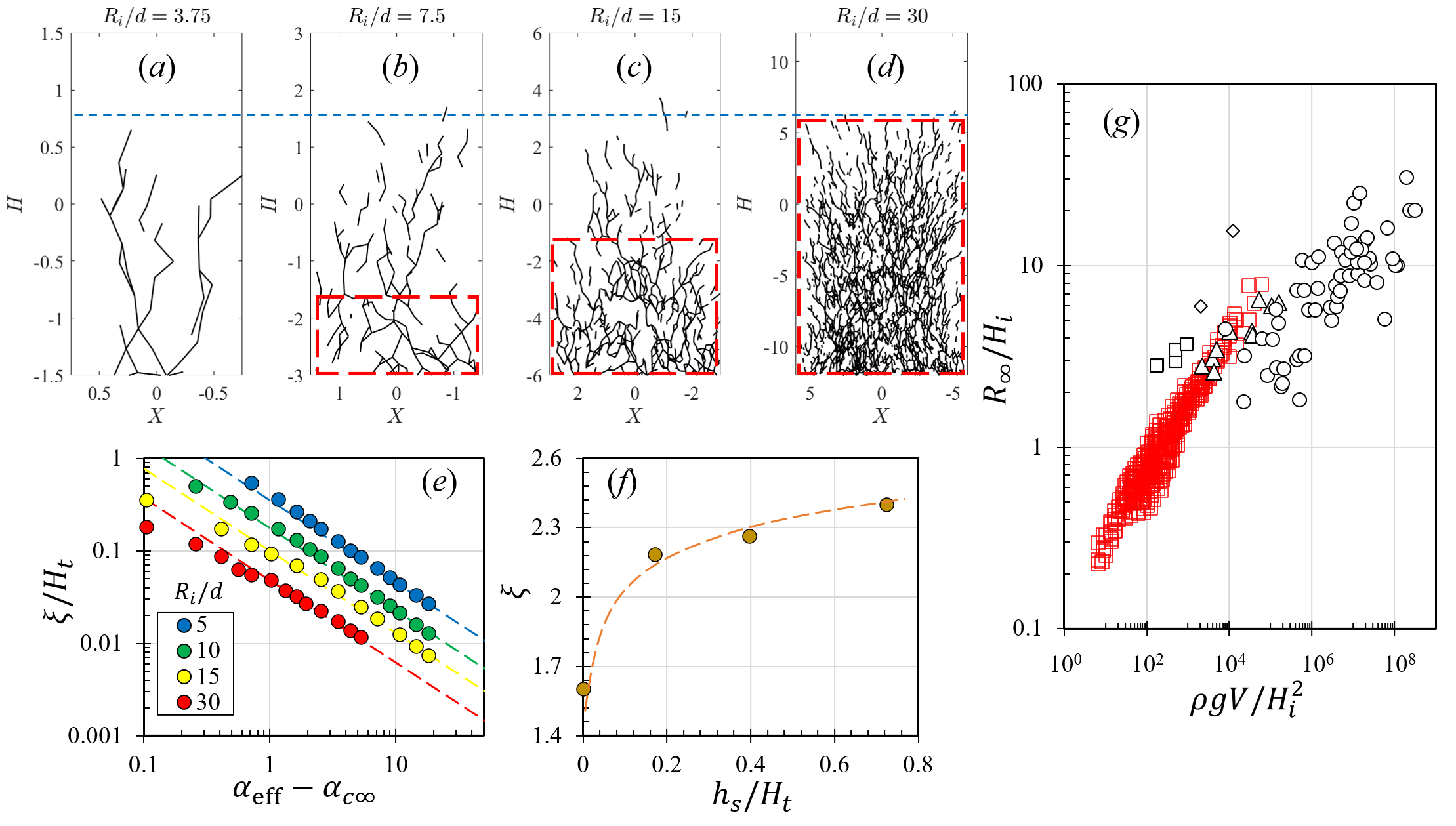}
  \caption{(a) - (d) show strong force networks of granular columns with the same initial aspect ratio but different relative sizes at the beginning of collapse. Solid lines in the figure represent the link between contacting particles with magnitude of contact forces larger than the mean magnitude of contact forces. The blue dashed line represents the height where the strong force network can propagate in the vertical direction, and red dashed lines denote the region where the strong force network could occupy the whole width. The height of the red dashed region is $h_s$, which represent the height of fully occupied strong force networks; (e) The relationship between $\xi/H_t$ and $\alpha_{\textrm{eff}} - \alpha_{c\infty}$ when $\xi$ and $H_t$ are measured at the beginning of the granular collapse, where $\xi$ is the correlation length scale, and $H_t$ is the height of the system at the time we take the measurement. We only measured $\xi$ for four different system sizes and plot the results using circle markers. Meanwhile, the fitted power-law relationships are denoted using dashed lines. (f) shows the relationship between $\xi$ and $h_s/H_t$, where $h_s$ is the height of the fully occupied strong force region in FIG. (a)-(d); (g) Compares simulation data to field data presented in \citeA{Calder1999MobilityOP}. The red \protect\tikzrect{red}{white} markers denote our simulation results, while other markers represent the data compiled from \citeA{Calder1999MobilityOP}.}
  \label{discussion}
\end{figure}

We then investigate the strong force network of columns with different relative sizes at the very beginning of collapses [Fig. \ref{discussion}(a)-(d)]. In Fig. \ref{discussion}(a)-(d), we only plot the contact forces $f$ that are larger than the mean contact force $\langle f\rangle$ linking the centroids of contacting particle pairs. The blue dashed lines represent the height that the strong force network can reach in the vertical direction, while the red dashed rectangles denote the region which the strong force network could occupy the whole cross-section. As we increase the relative size of the column, the height of the strong force network remains almost unchanged. However, the horizontal occupation of the strong force network is strengthened with the increase of the relative column size. This indicates that larger columns could form better-connected strong force networks in the horizontal direction at the beginning of the column collapse, and therefore might have the capacity to drive the system to flow further than smaller columns.

Equation \ref{eq_fss} and the strong force network imply that a length scale exists during the collapse of granular columns in a form that the relative size of the length scale with respect to the column height has a power-law relationship with $\alpha_{\textrm{eff}} - \alpha_{c\infty}$ of the form
\begin{linenomath*}
\begin{equation} \label{eq_length1}
    \begin{split}
        \xi/H_i \sim (\alpha_{\textrm{eff}} - \alpha_{c\infty})^{-\zeta},
    \end{split}
\end{equation}
\end{linenomath*}
where $\zeta$ is the critical exponent for the length scale $\xi$, and $\xi/H_i$ denotes the degree of occupation of sheared granular media along the height of the initial granular column. To further investigate the length scale $\xi$ associated with the collapse of granular columns, we choose four sets of simulations with different relative system sizes. We consider grain pairs along the vertical direction at the very beginning of the collapse when time $t =$ 0.05 s, and define the corresponding correlation $G(z)$, which describes the correlation between particle pairs with vertical distance $z$, based on the equation proposed by Barker and Mehta \cite{Barker1992,mehta2007granular}, by
\begin{linenomath*}
\begin{equation}
    \begin{split}
        G(z) = \frac{\langle\Delta{z_i}\Delta{z_j}\delta\left(|z_{ij}| - z\right)\Theta\left(|d_{ij} - 1/2|\right)\rangle}{\langle|\Delta{z_i}|\rangle^2}\ ,
    \end{split}
\end{equation}
\end{linenomath*}
where $\Delta{z_i}$ and $\Delta{z_j}$ are the vertical displacements of particle $i$ and $j$ after one time step (e.g. $\Delta{z_i} = v_i\Delta{t}$ where $v_i$ is the velocity of particle $i$ and $\Delta{t}$ is the time step), $z_{ij}$ is the vertical distance between two particles, $z$ is the vertical displacement between particles, and $d_{ij}=\sqrt{(x_i - x_j)^2 + (y_i - y_j)^2}$ is the horizontal distance between two particles. Here $\delta(\ )$ is a rectangular function, where the function is equal to one when $z-0.5d < |z_{ij}| \leq z+0.5d$ and equal to zero elsewhere, and $\Theta(\ )$ is the Heaviside step function. This definition ensures that the averages run over all displacements of sphere pairs in the vertical direction. The correlation function $G(z)$ shows how particles influence each other in the $z-$direction. 

The displacement correlation function can be fitted with an exponential equation $G(z) = A\textrm{exp}(-z/\xi)$, where $\xi$ can be seen as a correlation length scale associated with the displacement correlation among particles during the collapse of granular columns, and $A$ is a fitting parameter. The occupation of correlated grains across the height of a granular column, represented by $\xi/H_t$, where $H_t$ \remove{here} is the height of the column at the time when we measure the correlation length scale (since $\xi$ is measured at the very beginning of the collapse, $H_t \approx H_i$), indicates the collective motion of particles during a collapse of granular columns. Thus, we plot the relationship between $\xi/H_t$ and $\alpha_{\textrm{eff}} - \alpha_{c\infty}$ in Fig. \ref{discussion}(e), and determine that the ratio between correlation length scale and the system height shows power-law decay as we increase $\alpha_{\textrm{eff}} - \alpha_{c\infty}$, which is consistent with the finite-size scaling of both run-out behavior of granular column collapses, and $\zeta = 0.89$ in Eq. \ref{eq_length1} best fits the power-law decay. When $\alpha_{\textrm{eff}}$ approaches $\alpha_{c\infty}$, $\xi/H_t$ starts to deviate from the power-law relationship. This is because when the initial aspect ratio is too small, only several layers of particles are present along the height of a granular column. 

Additionally, we calculate the correlation length scales, $\xi$, of the four cases shown in Fig. \ref{discussion}(a)-(d), and plot them against the ratio between the height of the fully occupied strong force region $h_s$ and the column height $H_t$ [Fig. \ref{discussion}(f)]. We find that, as we increase the relative column size, not only does the height of the fully occupied strong force region increase, but the correlation length scale also increases accordingly. This implies that a larger system could help establish a better correlated moving mass, and as we increase the initial aspect ratio, columns with larger relative size, where well-correlated moving mass is easier to establish, would experience the phase transition from quasi-static collapse to inertial collapse earlier than smaller systems. To further analyze the phase transition associated with granular column collapses, we would have to conduct rigorous investigations on the mean-field theory, which we plan to do in future studies.

\subsection{Application to geophysical flows}
\label{sec:application}
In order to present an application of granular column collapse results to a real geological situation, we also compare our simulation data to the rock falls around Soufriere Hills Volcano, Montserrat as a result of volcanic eruptions and the resulting pyroclastic flows and granular surges [presented in \citeA{Calder1999MobilityOP}]. Although granular column collapses do not behave exactly the same as landslides, rock avalanches or pyroclastic flows, their underlying physics is similar since the mobility of collapsed materials is mainly governed by the ability of transforming potential energy into kinetic energy, by surpassing the dissipative action of friction, a competition of effects that is summarized in the proposed $\alpha_{\textrm{eff}}$ as explained before. \citeA{Roche2002ExperimentsOD,Roche2008ExperimentalOO} also argued that the dam-break experiments could reveal similar behavior as inertial geophysical flows and the results supported the pyroclastic flow modeling approach by \citeA{Levine1991HydraulicsOT}. In \citeA{Calder1999MobilityOP}, $R_\infty/H_i$, which indicates the ability of relocating materials, is plotted against a parameter, $\Psi = \rho gV/H_i^2$, where $V$ is the volume of the material being transported. In Fig. \ref{discussion}(g), we plot our simulation results along with the results presented in \citeA{Calder1999MobilityOP}, where the \tikzrect{red}{white} markers indicate our simulation results, while other markers represent the field data. 

We see that most of simulation data do not fall into the region of field observations. This is mainly due to the length scale of simulations being much smaller than a typical geophysical flow. However, as we increase the system size of simulations, we do obtain granular collapses in the same regime as some of the pyroclastic flows and granular surges. The finite-size analysis early in this section implies that increasing the system size could lead to an asymptotic behavior of granular flows, where the transition point $\alpha_c \rightarrow \alpha_{c\infty}$. We can also see this trend in Fig. \ref{discussion}(g) where, as we increase the system size of simulations, our simulation data merge into the field results. More intriguingly, this also shows that almost all the data acquired from dome-collapse pyroclastic flows [\tikztri{black}{white} in Fig. \ref{discussion}(g)] fit in the trend of simulation results; this type of pyroclastic flows is considered to have a flow behavior more similar to granular flows. The size effect of granular column collapses might be potentially connected to predict the behavior of geophysical flows with different length scales. More importantly, the scaling of run-out distances with respect to $R_i/d$ might help explaining the long-range run-out in some pyroclastic flows \cite{Lube2019GenerationOA}, where the initial radius of the pyroclastic ``column" is much larger than the volcanic particles. The analysis of force networks within granular systems, as mentioned in Section \ref{sec:size}, may be helpful to explain some phenomena of geophysical flows. However, it is still difficult to directly obtain the contact information within real geophysical systems.

\section{\label{Conclusion}Conclusions}

Previous research concluded that we could combine the influence of initial and boundary conditions and the initial column aspect ratio to determine the normalized run-out distance with an effective aspect ratio, $\alpha_{\textrm{eff}}$, which introduced a universal relationship to link the behavior of granular column collapses in three different collapsing regimes (quasi-static, inertial, and liquid-like) \cite{man2020universality}. In this paper, we further investigated the size effect associated with the collapse of granular columns. Our research is performed with DEM simulations of Voronoi-based spheropolyhedron particles. We found that the transition point in $\mathcal{R}-\alpha_{\textrm{eff}}$ space, which distinguishes the inertial collapse regime from the quasi-static collapse regime, varies as we change the relative system size $R_i/d$. Both $\mathcal{R}_c$ and $\alpha_c$ experience power-law decay with respect to $R_i/d$, which implies possible finite-size scaling for the normalized run-out distance $\mathcal{R}$.

Similar to the finite-size analysis of the jamming transition of granular materials \cite{Liu2014finite}, where the stress scales with the solid fraction $\phi - \phi_{c\infty}$, we took the previously determined effective aspect ratio $\alpha_{\textrm{eff}}$ as the key parameter, and discovered that both the run-out distance of granular column collapse and the energy consumption of it follows strong finite-size scaling. Interestingly, the scaling parameters we discovered for the finite-size scaling of the run-out distance of granular column collapses are similar to those presented in percolation problems of two-dimensional random networks \cite{Torres2015scaling}, which may imply that the behavior of granular column collapses is strongly influenced by the contact network presented inside the column during the collapse. Additionally, to better understand the scaling of granular column collapses, we further analyze the correlation length scale at the very beginning of the collapse. Simulation results show that, as we increase $\alpha_{\textrm{eff}} - \alpha_{c\infty}$, the length scale $\xi/H_t$ shows a power-law decay with the critical exponent $\zeta\approx0.89$. This study is based on our previous work where we introduced a physics-based dimensionless number (ratio between inertial effects and frictional effects) to describe the behavior of granular column collapses, and we further expand our analysis to include the size effect, which is crucial to applications in engineering, such as chemical engineering, food processing, civil engineering. We also associate the size effect to the strong force networks at the beginning of the column collapse to show that a larger column size results in a well percolated strong force network, which may contribute to relatively larger run-out distances. 

However, the granular system we are studying is still axisymmetric, and our preliminary results show intriguing phenomena when the cross-section of a granular column is no longer axisymmetric. Also, we note that the system we are dealing with has no interstitial fluid, and only Voronoi-based particles are considered. In natural and engineering applications, most granular systems are wet, saturated, or submerged in liquid. A rigorous investigation on the fluid-solid interaction and its influence on the bulk behavior, which is beyond the scope of this study, is need to thoroughly describe the behavior of granular-fluid systems. Further investigations related to granular-fluid systems, the influence of cross-section shapes, and the influence of particle shapes will be conducted and presented in future publications.

\section*{Data availability Statement}
The data associated to this work is archived and published as \citeA{Man2021data} in Mendeley Data, which can be accessed freely after registration. Part of the data in Fig. \ref{discussion}(g) is available through \citeA{Calder1999MobilityOP}. The simulation and the corresponding model were described in \citeA{Mechsys} and \citeA{galindo2013coupled}. The software can be obtained from \url{http://mechsys.nongnu.org/}.

\acknowledgments
The authors acknowledge the financial support from Westlake University and thank the Westlake University Supercomputer Center for computational resources and related assistance. H.E. Huppert acknowledges with gratitude the hospitality of his co-authors while he was at Westlake University.


%
%

\bibliography{GranCollapse_SizeEffect}

%
%
%
%
%

\end{document}


%
%


\title{Supporting Information for "Finite-Size Analysis of the Collapse of Dry Granular Columns"}
%
%

%
%



\authors{Teng Man\affil{1}, Herbert E. Huppert\affil{2}, Ling Li\affil{1}, Sergio A. Galindo-Torres\affil{1}}

\affiliation{1}{Key Laboratory of Coastal Environment and Resources of Zhejiang Province (KLaCER), School of Engineering, Westlake University, 18 Shilongshan Street, Hangzhou, Zhejiang 310024, China}
\affiliation{2}{Institute of Theoretical Geophysics, King's College, University of Cambridge, King's Parade, Cambridge CB2 1ST, United Kingdom}

%
%

%

\begin{article}

%
%

\noindent\textbf{Additional Supporting Information (Files uploaded separately)}
\begin{enumerate}
\item Data set for the size effect of granular column collapses
\item Movie S1 of a granular column collapse
\item Movie S2 of a granular column collapse
\end{enumerate}

\noindent\textbf{Introduction}
In this supplemental material, we provide the data we obtained from the DEM simulation and used to generate figures. Additionally, we provide two movies of the simulation of collapse of granular column.



\noindent\textbf{Data set for the size effect of granular column collapses} 


\noindent\textbf{Movie S1 of a granular column collapse} 


\noindent\textbf{Movie S2 of a granular column collapse} 
\end{article}
\clearpage